\documentclass[aps, prb, twocolumn,showkeys,showpacs,amsmath,amssymb,superscriptaddress]{revtex4-1}

\usepackage[T1]{fontenc}
\usepackage[latin1]{inputenc}

\usepackage[ngerman,british]{babel}
	\addto\captionsngerman{}

\usepackage{amsmath,amssymb,amsthm,dsfont,mathtools,mathrsfs,bbm}

\usepackage{mdwlist}
\usepackage{paralist}
\usepackage{array}
\usepackage{booktabs}						
\usepackage{dcolumn} 						

\usepackage{graphicx}
\usepackage{wrapfig}

\usepackage{url}

\usepackage{color}
	\definecolor{myblue}{rgb}{0,0.3,0.8}
	\definecolor{mygreen}{rgb}{0,0.5,0}
	\definecolor{myblue}{rgb}{0,0.3,0.8}

\usepackage[version=3]{mhchem}

\usepackage{savesym}

\savesymbol{a}
\savesymbol{b}
\savesymbol{d}
\savesymbol{e}
\savesymbol{k}
\savesymbol{l}
\savesymbol{H}
\savesymbol{L}
\savesymbol{p}
\savesymbol{r}
\savesymbol{s}
\savesymbol{S}
\savesymbol{F}
\savesymbol{t}
\savesymbol{w}
\savesymbol{th}
\savesymbol{div}
\savesymbol{tr}
\savesymbol{Re}
\savesymbol{Im}
\savesymbol{AA}
\savesymbol{SS}
\savesymbol{Si}
\savesymbol{gg}
\savesymbol{ss}

\newcommand{\a}{\alpha}
\newcommand{\b}{\beta}
\newcommand{\e}{\varepsilon}
\newcommand{\d}{\delta}
\newcommand{\del}{\partial}

\newcommand{\g}{\gamma}
\newcommand{\G}{\Gamma}

\newcommand{\s}{\sigma}

\newcommand{\w}{\omega}

\newcommand{\dt}{{\Delta t}}

\newcommand{\cov}{{\nabla}}

\newcommand{\C}{\mathbbm{C}}

\DeclareMathOperator{\Im}{Im}

\DeclareMathOperator{\tr}{Tr}

\newcommand{\T}{\textstyle}

\newlength{\tmplen}

\usepackage{bbold}

\begin{document}

\title[Moebius graphene strip]{Curvature-induced quantum spin-Hall effect on a M{\"o}bius strip}

%

%
%
%
\author{Kyriakos Flouris}
\author{Miller Mendoza Jimenez}
\affiliation{ %
ETH
  Z\"urich, Computational Physics for Engineering Materials, Institute
 for Building Materials, Wolfgang-Pauli-Str. 27, HIT, CH-8093 Z\"urich
 (Switzerland)}%

\author{Hans J. Herrmann}
\affiliation{ %
Departamento de F\' isica, Universidade do Cear\' a, 60451-970 Fortaleza, Brazil}%
\affiliation{ %
C.N.R.S., UMR 7636, PMMH, ESPCI, 10 rue Vauquelin, 75231 Paris Cedex 05, France }%

\begin{abstract}%




The quantum Hall effect has been predicted and discovered in various condensed matter systems. A promising quantum material for such topological effects is graphene. 
We report the numerical observation of a curvature induced spin-Hall effect in a monolayer graphene M\"obius strip. The solution of the Dirac equation on the non-trivial and non-Eucleadean manifold reveals that a despite the absence of a Hall current, a spin-Hall current is a natural consequence for such a topology as predicted from symmetry arguments.

\end{abstract}

\maketitle

\section{Introduction}
The discovery of the quantum Hall effect has triggered a surge of research in topological states of matter with a multitude of applications in quantum engineering and condensed matter physics  \cite{haldane_review, topo_insulators_1, topologicalstatesofmatter}. The topological invariants \cite{chernnumber} of a system provide simplicity in the implementation, as well as, protection from disorder \cite{topo_disorter}. Naturally, graphene has been extensively investigated in this context, inheriting its remarkable characteristics \cite{graphenerev2, graphenerev1, graphenerev3} into a topologically protected system. For example, the honeycomb bipartite lattice of graphene has been shown to exhibit the quantum Hall and the quantum spin-Hall effects in the presence of a magnetic field \cite{graphene_Hall, graphene_spin_Hall, graphene_haldane}. Nevertheless, a much more promising system would require a breaking of the time invariance in a more intrinsic way without the need of an external field \cite{edge-state_withou_hflield,qhe_without_hfield}.

 One such system is the simplest non-trivial fiber bundle, the famous M\"obius strip \cite{calvao, polymer}.  Theoretical analysis has already hypothesized that a  graphene M\"obius strip can exhibit topological insulator properties owning to the zig zag edge states \cite{mobius_strip, ci_qah}.  In this paper we show that a graphene M\"obius strip inherently exhibits a quantum spin-Hall current. The strains induced in curved graphene are modeled via a non-Euclidean space description. The result is alluring, because a pure, curved space formulation of the lower energetic states of graphene would inevitably create a spin-Hall current and not a Hall current as theoretically predicted via symmetry arguments \cite{Nontrivial}. Furthemore, curvature has immense potential as a novel and straightforward control mechanism in quantum devices \cite{molfetta2013, Cortijo2007}.

 In this paper, we carry out the a numerical study of the quantum spin Hall effect on the curved space solution of a graphene M\"obius strip. We simulate Dirac particles and relate them to graphene through the low energy approximation of the band structure. The effect of curvature on transport and the energy levels of confined Dirac particles on such a manifold are investigated.  The non-relativistic quantum mechanics on a the M\"obius strip has been studied  \cite{mobius_graphene_QM} and non-trivial effects on the M\"obius topology have also been proposed for molecular devices \cite{mobius_molecular_devices}. It has already been shown that in graphene, the quantum Hall effect can be realized with careful strain engineering in the zero field case \cite{zerofield_qhf}. 
 
  To this end  numerical studies were curried out with a solver of the Dirac equation in curved space \cite{JD_thesis, cdp}. The method is based on the conceptual similarities between the Dirac and Boltzmann equations and is an extension of the quantum lattice Boltzmann  method  \cite{succi_qlbm} to curved space. Through our simulations we observe a quantum spin-Hall current in the bulk of a M\"obius graphene strip. Additionally, we compare the result to a simpler curved space topology, namely the torus and we propose a simple valid current conserving and time-reversal symmetric boundary condition for the method. Furthemore, a specific illustration of the equivalence between the Berry and Ricci curvature is presented analytically through a traveling wave-packet around a M\"obius strip.

\section{The Dirac equation in curved space}

 By minimally coupling the Dirac equation to curved space and a vector potential, $A_i(x)$, the Dirac Hamiltonian takes the form:
 \begin{equation}
\label{eq:hamiltoniandirac}
H_D=-i \int \Psi^\dagger \sigma^a e_a^{~i}( \del_i + \G_i-i A_i)\Psi \sqrt{g} d^2x,
\end{equation}
 in natural units such that $\hbar=c=1$ where $\hbar$, is Planck's constant and $c$, the speed of light, here $m$ is the particle mass, $\mu={0,1,2}$ for two-dimensional space-time. Here $\Psi = (\Psi^+, \Psi^-) = (\psi_1^+,\psi_2^-,\psi_1^-,\psi_2^+) \in \C^4$ denotes the spinor, and $\g^\mu = \g^\a e_\a^{\ \mu}$ are the generalized $\g$-matrices, where $\g^\a \in \C^{4\times 4}$ are the standard $\g$-matrices (in Dirac representation). $e_\a^{\ \mu} $ is the tetrad (first index: flat Minkowski, second index: curved space-time). Here, the tetrad is defined by $e_{\alpha}^{\mu} g_{\mu \nu}e_{\beta}^{\nu}=\eta_{\alpha \beta}$,
where $g_{\mu \nu}$ denotes the metric tensor and $\eta_{\alpha \beta}$ is the 
Minkowski metric. The tetrad basis is chosen such that the standard Dirac matrices can be utilized with no need to transform to a new coordinate basis. 
$\G_{\mu}$ denotes the spin connection matrices given by $\Gamma_\mu = - \frac{i}{4} \w_\mu^{\a\b} \s_{\a\b},$ where $\s_{\a\b} = \frac{i}{2} [\g_\a,\g_\b]$ and $\w_\mu^{\a\b} = \quad e_\nu^\a \cov_\mu e^{\nu \b}$.
The Dirac equation in curved space describes quantum relativistic Dirac particles (e.g., electrons ) 
moving on arbitrary manifold~trajectories. 

To model the single layer carbon atom honeycomb lattice structure we start from the tight binding Hamiltonian which is constructed assuming superposition of local wavefunctions for isolated atoms on a honeycomb lattice \cite{graphene_tb1}. In the low energy limit it has been shown that the tight binding Hamiltonian converges to the Dirac Hamiltonian in the continuum limit. Therefore, for graphene the effective Hamiltonian is \cite{OLIVALEYVA}
\begin{equation}
\label{eq:hamiltoniangraphene}
H^*_D=-i v_f \int \Psi^\dagger \sigma^a (v_a^{*i} \del_i + \G_a^*-i A^*_a)\Psi d^2x,
\end{equation}
where $v_a^{* i}=\d_{a i} + u_{a i} -\beta \e_{a i}$ is the space dependent Fermi velocity, $\G^*_a=\frac{1}{2 v_f}\del_j v_a^{* j}$ is a complex gauge vector field which guarantees the Hermiticity of the Hamiltonian and $A^*_a$ is a strain-induced pseudo-vector potential given by $A^*_a=(A_x^*,A_y^*)=\frac{\beta}{2a}(\e_{xx}-\e_{yy},-2\e_{xy}$), $\beta$ is the material dependent electron Grueneisen parameter, $a$ the lattice spacing and $\e_{i\jmath}= u_{i\jmath} +\frac{1}{2}\del_i h \del_j h$ the general strain tensor with in-plane, $u_{i\jmath}$ and out of plane, $h$ deformations. The term $u_{a i}$ in $v_a^{* i}$ can be interpreted as the deformation potential and is purely a geometric consequence due to lattice distortion, it does not depend on the material as long as it has the same topology. Comparing this to the standard Dirac Hamiltonian in curved space (\ref{eq:hamiltoniandirac}) we can match both Hamiltonians $H_D$ and $H^*_D$. The numerical solutions are obtained with the quantum lattice Boltzmann method as described in  Appendix and Ref.~\cite{debus_qlb}.

\section{Method and Results \label{sec:moebius_strip}}

\subsection{Quantum spin-Hall effect on a M\"obius strip}
\subsubsection{Geometry and boundary conditions}

The system geometry is initialized to the M{\"o}bius strip by the discrete mapping (or chart),
  \begin{align}
    h^\alpha(\theta,r)	= & \begin{pmatrix}
           [R+ w r \cos(\eta\theta/2)]cos(\theta) \\
           [R+ w r \cos(\eta\theta/2)]sin(\theta) \\
           w r \sin(\eta\theta/2)
         \end{pmatrix}
  \end{align}
with $\theta \in \{-\pi, \pi \}$, $ r \in \{-L_r/2, L_r/2 \}$,  half-width $w$, mid-circle of radius $R$, and number of turns $\eta$ at height $z=0$. $L_r$ is the domain size across the radial direction, for simplicity we set $L_r/2=1$. In the simulations, we consider a square sheet with reverse periodic boundary conditions in one direction on a grid of size $l_\theta \times l_r= 512 \times 512$ or 100 $\times$ 100 nm$^2$, $A_a$; the external potential is set to zero.

The metric tensor can be computed from the $h^\alpha(x,y) $ relating the positions of the atoms from the three dimensional flat space (laboratory frame with Minkowski-metric) to the curved space by:
\begin{equation}
    g_{ij}= \frac{\del h^\alpha(\theta,r)}{\del x^i}\frac{\del h^\beta(\theta,r)}{\del x^j}\eta_{\alpha \beta},
\end{equation}
With this parametrisation the metric is cast to a diagonal form,
\begin{align*}
	g_{i j} = 
	\begin{pmatrix}
     	G_{11}^2 & 0 \\ 
		0 & G_{22}^2	
	\end{pmatrix}.
\end{align*}
where $G_{11}=\sqrt{R+wr\cos(\eta\theta / 2)+w^2r^2/4}$ and $G_{22}=w$. The reversed periodic boundary at $\theta=\pi$ is implemented as $\Psi(\theta=-\pi,r)=\Psi(\theta=\pi,-r)$. From Ref.~\cite{boundary_graphene} a valid current conserving and time-reversal symmetric boundary condition for the Dirac equation can be written as:
\begin{equation}
\Psi=(\mathbf{v} \cdot \mathbf{\tau}) \otimes (\mathbf{n} \cdot \mathbf{\sigma}), ~ \mathbf{n} \bot \mathbf{n}_B,
\end{equation}
where $\mathbf{n}_B$ is the unit vector in the $r-\theta$ plane normal to the boundary, $\mathbf{v}$ and $\mathbf{n}$ are three dimensional unit vectors  and $\mathbf{\tau}$, $\mathbf{\sigma}$ are Pauli matrices. In the case of a mass-less  particle inside a one-dimensional box, assuming perpendicular reflection, it is sufficient to set $\Psi_{1,2}(r=-1)=\Psi_{1,2}(r=1)=0$. This closed boundary condition ensures probability conservation for the decoupled Weyl equations such that $\rho(r=-1)=\rho(r=1)$ and $J^\mu(r=-1)=J^\mu(r=1)=0$. Numerically, for the present simulational timescales, we observe a small $(<1\%)$, resolution convergent error related to the finite size effect of the wavefunction. The error convergence plot is shown in Appendix \ref{app:boundary}.

\begin{figure}
\includegraphics[width=1\columnwidth]{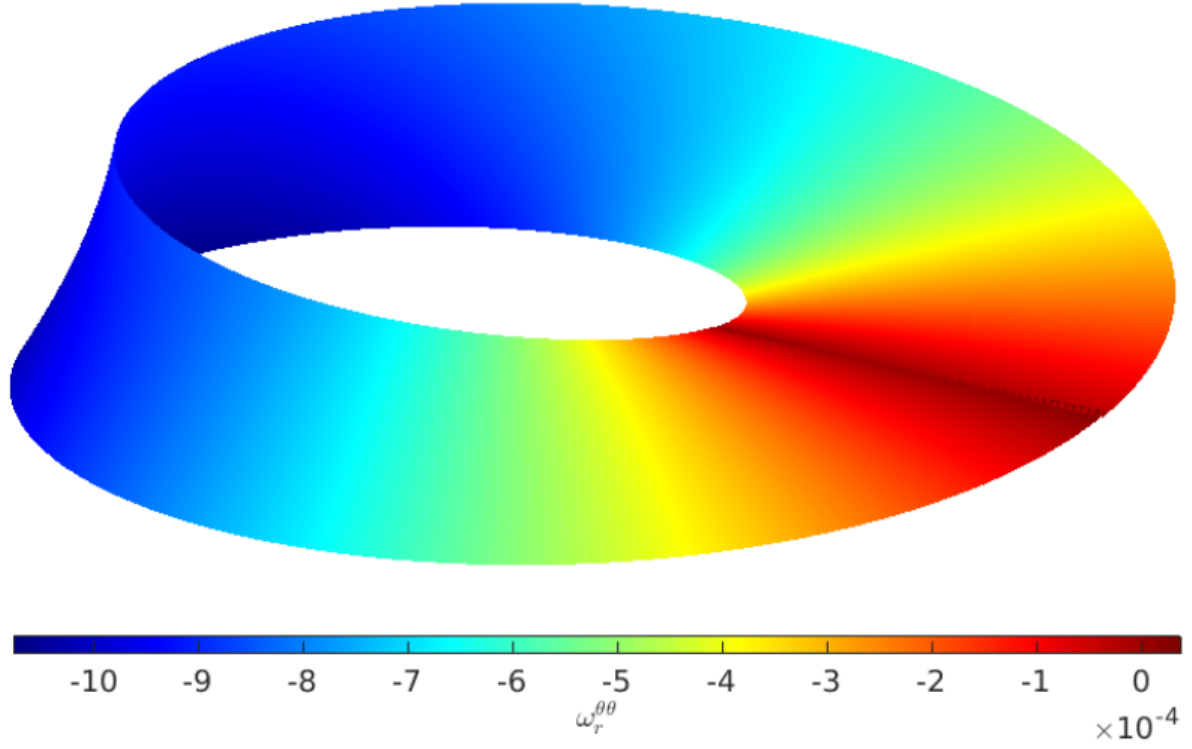}
\caption{\label{fig:spin_connection} The spin connection component on the M\"obius strip for half-width $w=0.1$ and mid-circle of radius $R=1$. }
\end{figure}

 The metric tensor, although diagonal, for the typical choice of $w\sim 1$ imposes some large gradients on the spin connection $\Gamma^i$ which introduce numerical instabilities. From the form of the metric it can be shown that for $w \ll 1$, the metric variation is minimized resulting in more stability, physically this results in a long and thin strip. It should be noted that this choice of parametrization results in non-zero curvature across the whole domain, expressed by the Ricci scalar, see the Appendix. Additionally, $\theta$ is intentionally chosen such that the spin connection is continuous across the reverse periodic boundary, see Fig.~\ref{fig:spin_connection}.  This is not the case for the most common convention $\theta \in \{0, 2\pi \}$. The spin connection (or the Christoffel symbols) is not a gauge independent quantity but the metric and Ricci tensors are. In fact, the complete continuity of the spin connection is achieved only for $w \ll 1$, for $w \sim 1$ there is still some discrepancy as $\omega_i^{jk}$ is not symmetric in $r$.

\subsubsection{Berry connection and symmetry arguments \label{app:berry_phase}}

Topological currents are traditionally described as being a consequence of the integral of the phase space connection, the Berry phase \cite{berry_paper}. Analogous to the real space curvature, the Berry connection is not gauge independent but the Berry phase and the Berry curvature are. The similarities of the Ricci and Berry curvatures are further investigated here, we show that a Gaussian wave-packet completing a circle around a M\"obius strip will attain a phase $\pi$ equivalent to a Berry phase. This is a consequence of the topology and represents a specific illustration of the relation between Berry and real space curvature \cite{QAC}.

To solve the Dirac equation, minimally coupled to curvature,
\begin{equation}
\label{eq:Dirac_}
(i \g^{\mu}D_{\mu})\Psi=0,
\end{equation}
where $D_{\mu}$ is the covariant derivative as $D_{\mu} \Psi=\del_{\mu} \Psi + \G_{\mu} \Psi$ we  assume that the wave-packet has a negligible profile, $\delta \mathbf{r} \rightarrow 0$.  The connection component of the covariant derivative can be absorbed into the wavefunction such that
\begin{equation}
    \Psi \rightarrow \Psi \exp\bigg(i \int_{\mathbf{r}_c}^{\mathbf{r}_c+\delta \mathbf{r}} \Gamma_i d\mathbf{r} \bigg) 
\end{equation}
where $\mathbf{r}_c$ is the center of mass position and $\Gamma_i$ is the spin connection matrix. For a Gaussian wavepacket with spread $\sigma$ and momentum $\mathbf{k}$
\begin{equation}
     \Psi(\mathbf{r},\mathbf{k})   = \frac{1}{\sqrt{2 \pi \mathcal{\s}^2}} 
     \begin{pmatrix}
           1 \\
           0 \\
           0 \\
           -1
         \end{pmatrix}
     e^{i \int \Gamma_i d\mathbf{r}} e^{-\frac{|\mathbf{r}|^2}{4\mathcal{\s}^2}+i\mathbf{k}\cdot\mathbf{r}}.
\end{equation}
 This wave-function effectively minimally couples the standard Dirac equation to curved space through the spin connection. We define the Berry connection as 
\begin{equation}
\label{eq:berry_conection}
    \mathcal{A}^i_n(\mathbf{R})=i\langle \Psi(\mathbf{R})| \del_\mathbf{R} | \Psi(\mathbf{R}) \rangle
\end{equation}
for some parameter space $\mathbf{R}$ and eigen-function $n$. The Berry phase can be calculated from the complete loop integral of the connection
\begin{equation}
    \gamma=\oint_0^{2 \pi} \mathcal{A}(\mathbf{R})g^{i j}_{\mathbf{R}}d\mathbf{R}.
\end{equation}
In a manner similar to the treatment of the Aharonov-Bohm effect from the Berry connection \cite{berry_paper}, we define the fast and slow coordinates as $R$ and $r$ respectively such that $\Psi(R) \rightarrow \Psi(r-R)$. 
For the M\"obius strip, choosing the real space to be the parameter of integration and  restricting the motion to one-dimension $r \in{0,2\pi}$, the center of the band, the wave-function takes the form
\begin{equation}
     \Psi_r(R-r)   = \frac{1}{\sqrt{2 \pi \mathcal{\s}^2}} 
     \begin{pmatrix}
           1 \\
           0 \\
           0 \\
           -1
         \end{pmatrix}
     e^{i \int \Gamma_r d r} e^{-\frac{|R-r|^2}{4\mathcal{\s}^2}+ik(R-r)}.
\end{equation}
From Eq.~\ref{eq:berry_conection} the explicit form of the wavefunction implies that $\mathcal{A}^i=\tr \Gamma_i$. The implication of these result is that the Berry connection and curvatures can be directly related to the real space affine connection and Ricci curvature tensor under some conditions.   

As a consequence, the phase change of a wavepacket moving around a M\"obius strip can be calculated from the Berry phase. Integrating naively around the band 
\begin{equation}
    \gamma=\oint_0^{2 \pi} \tr \langle \Psi_r| \del_r  \Psi_r \rangle g^{1 1} d\mathbf{r},
\end{equation}
where $\tr$ denotes the trace of the resulting matrix and it takes into account the spinor nature of the Dirac wavefunction, yields a trivial result: $\gamma=0$. The caveat is that, in this coordinate basis, for both half-width and half-radius equal to unity,  $\Gamma_r$  simplifies to a diagonal matrix such that 
\begin{equation}
\left(
\begin{array}{cccc}
 \frac{1}{2} i \cos \left(\frac{r}{2}\right) & 0 & 0 & 0 \\
 0 & -\frac{1}{2} i \cos \left(\frac{r}{2}\right) & 0 & 0 \\
 0 & 0 & \frac{1}{2} i \cos \left(\frac{r}{2}\right) & 0 \\
 0 & 0 & 0 & -\frac{1}{2} i \cos \left(\frac{r}{2}\right).
\end{array}
\right). 
\end{equation}
$\Gamma_r$ is discontinuous when $\theta=2\pi \rightarrow 0$. To make the function single valued we can perform a gauge transformation such that $\Psi \rightarrow \Psi'=\Psi e^{i\frac{r}{2}}$. This implies that $\gamma' = \gamma - \oint dr/2=-\pi$. Therefore the wavefunction picks up a phase of $\pi$ as it moves around the band.
This solutions clarifies the connection between the phase and real space curvatures and emphasizes that a topological current is expected from geometrically non-trivial manifolds.

In the case of low energy graphene, the calculated phase will be equal and opposite between the $k<0$ and $k>0$ states for a wavepacket on the central line. To observe a non-trivial current for a graphene strip we need to depart from the center-line approximation and the edges should be taken into consideration.   Furthermore since the phase will oscillate the total Hall currents are expected to be zero, which is not the case for the spin-Hall currents, as explained also in Beugeling et al. \cite{Nontrivial}. The argument follows from the non-orientability of the M\"obius strip implying that any pseudovectorial field can not be defined globally and smoothly on such a surface. The obvious consequence of this is the definition of a Chern number within such  topology. The introduction of spin $1/2$ degrees of freedom creates a unique orientable manifold that has an injective map to the original (base) space, i.e. an orientable double cover (ODC). Subjecting this ODC to a Haldane-like flux, by virtue of the geometry alone, two Haldane fluxes  of opposite chirality are expected to develop for the particle-hole pair.  This would be an intrinsic SO coupling inducing spin-Hall currents; that is, the two edge modes will counter-propagate.

\subsubsection{Numerical results}

To study the topological properties and observe a Hall-type effect it is necessary to realize a forward moving wavefunction that explores the complete domain. An electric field is both experimentally challenging and theoretically inconsistent across a periodic system such as a M\"obius graphene strip. Additionally, a magnetic field is difficult to keep tangential to such a manifold and unnecessary, as shown later. Consequently, a wave function with non-zero $k_\theta$ is initialized in the domain for a particle and a hole respectively,
\begin{align}
    \psi_P(\mathbf{k},\mathbf{r})=  & \begin{pmatrix}
           i \sin(k_\theta \theta) \\
           0 \\
           0 \\
          \frac{\hbar v_f \mathbf{k} }{E}  \cos(k_\theta \theta)
         \end{pmatrix}
         ~
    \psi_H(\mathbf{k},\mathbf{r})=  & \begin{pmatrix}
           0 \\
            \frac{\hbar v_f \mathbf{k} }{E} \cos(k_\theta \theta) \\
            i \sin(k_\theta \theta)\\
           0 
         \end{pmatrix}.     
  \end{align}
where $\hbar=v_f=1$, $E=|k|$, $\mathbf{r}=(\theta, r)$ and  $\mathbf{k}=(k_\theta,k_r),~k_r=0$.

\begin{figure}
\includegraphics[width=\columnwidth]{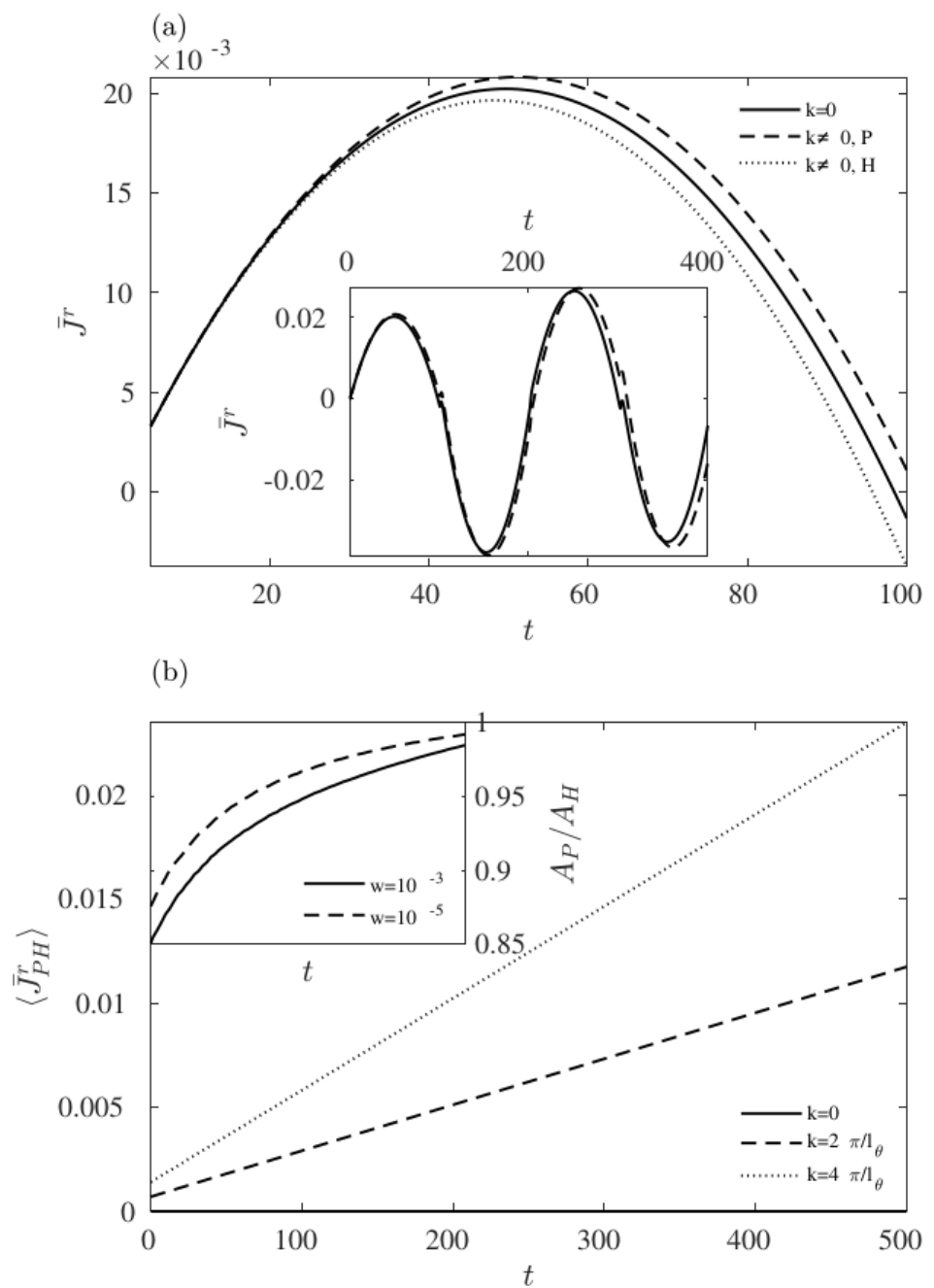}
\caption{\label{fig:spin_Hall} \textbf{(a)} The time evolution of the radial space-integrated Dirac current  $\bar{J}^r$ for the zero-momentum particle wavefunction and non-zero momentum particle and antiparticle wave-functions. The inset shows the time evolution of $\bar{J}^r$  for a particle wavefunction for zero and non-zero momentum.  \textbf{(b)}  The time evolution of the average particle-hole space-integrated current for zero and two different non-zero $k_\theta$. The inset shows the asymmetry ratio for two different half-widths $w$. One simulational time-step, t, is equivalent to $1ps$.}
\end{figure}

In  this section the subscript $\theta$ is dropped for brevity, $k_\theta \rightarrow k$. The time evolution of the radial space-integrated Dirac current
\begin{equation}
\bar{J}^r= \bigg( \int\limits^{x,y} J^r(x,y) \sqrt{g}dx dy\bigg)  
\end{equation}
for $\psi_P$ is shown in  Fig.~\ref{fig:spin_Hall}(a); for the zero-momentum and $k=2\pi/l_\theta$ cases, for both the particle and hole wave-functions. The dashed  and the dotted lines lie on opposite sides relative to the $k_\theta=0$ scenario, indicating a net anomalous velocity effect (in the $r$-direction) as a consequence of a $\theta$ velocity, i.e. a Hall current. The time evolution of $\bar{J}^r_{P}$ and  $\bar{J}^r_{P,k=0}$ for  longer times  are plotted in the inset of Fig.~\ref{fig:spin_Hall}(a). The oscillations are a result of the geometry and the closed boundary condition because they are also present for $\bar{J}^r_{P,k=0}$. There is no obvious non-zero average $\bar{J}^r_{P}$, or quantum Hall current.

The asymmetry ratio, $A_P/A_H$, is plotted in the inset of Fig.~\ref{fig:spin_Hall}(b) for two different half-widths $w$, defined by
\begin{equation}
    \frac{A_P}{A_H}=\frac{\bar{J}^r_{P}-\bar{J}^{r}_{P,k=0}}{\bar{J}^{r}_{H}-\bar{J}^{r}_{H,k=0}},
\end{equation}
where $\bar{J}^r_{P,H}$ denote the particle-hole currents for $k=2\pi/l_\theta$ and $\bar{J}^r_{P,H,k=0}$ denote the particle-hole currents for  $k=0$. $\bar{J}^r_{P,k=0}$ and $\bar{J}^r_{H,k=0}$ are equal to each other. The dependence of the asymmetry ratio on $w$ suggests that the difference between $\bar{J}^r_{P}$ and $\bar{J}^r_{H}$ is a consequence of  the curvature of the manifold and implies again a net anomalous velocity effect.

In Fig.~\ref{fig:spin_Hall}(b) the time evolution of the average particle-hole space-integrated current $\langle \bar{J}^r_{PH} \rangle= |\bar{J}^r_P-\bar{J}^r_H|$, is plotted for two different non-zero $k$. The result clearly shows a quantum spin-Hall current (or particle- antiparticle current), which is $k_\theta$ dependent. The effect is a transient result as there is no constant electric field in order to measure a stationary solution. Since a M\"obius strip has only one boundary, it is easier to visualize the result when it is projected to a rectangle (simulational domain), where the current is flowing towards one direction.

The spin-Hall current is a consequence of the curvature and topology of the system. The gauge field breaking the symmetry of the system is the spin connection matrix $\Gamma_i$, similar to a twisting magnetic field. For graphene, these curvature effects are realized as forces due to the pseudo-potential in a strained sheet \cite{OLIVALEYVA}. The correspondence between a magnetic field and curvature is also evident in Sec.~\ref{app:berry_phase}, as the Berry phase is directly related to the Aharanov-Bohm effect \cite{aharanov-bohm}. 

The space integrated spin-Hall current is then a result of the topology, i.e. the twist. This can be understood theoretically as a space-integrated current would not be affected by the 'sharpness' of the twist and any 'flat' regions are irrelevant. Moreover, assuming geometrical disorder is not stronger than the cumulative effect of the twist,  the bulk current will not be affected more than a perturbative correction. 

\begin{figure}
\includegraphics[width=\columnwidth]{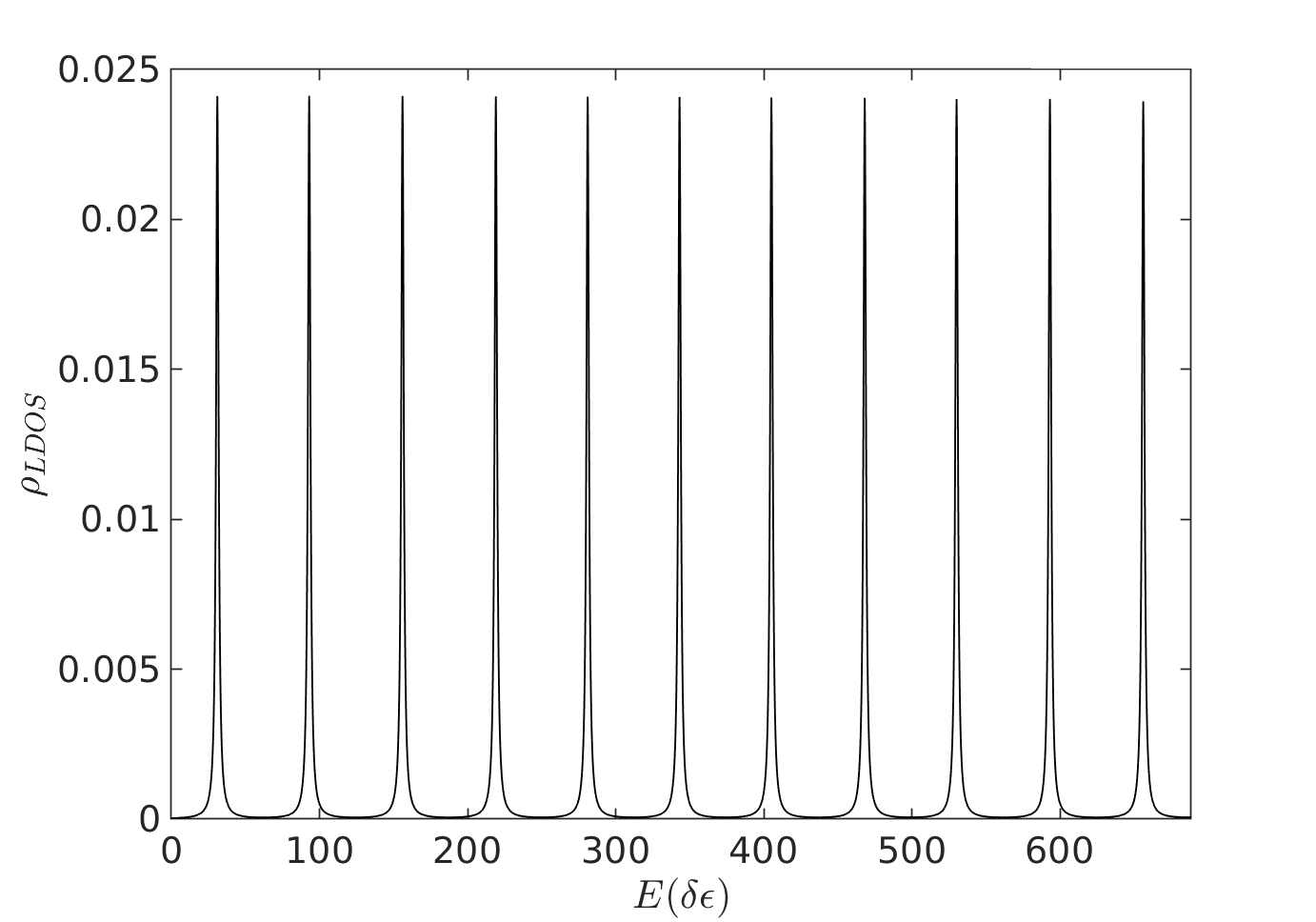}
\caption{\label{fig:LDOS}\textbf{} Local density of states plotted against normalised energy at   $\theta=r=0$.}
\end{figure}

Furthermore,  it has been established that the quantum levels on the M\"obius strip are quantized in $k_\theta$ for the Schr\"odinger equation, see Ref.~\cite{mobius_graphene_QM}. Equivalently this quantisation is expected also for the Dirac equation indeed this is evident from the local density of states (LDOS) as calculated from the energy spectrum $E_n$ of the system and its normalized eigenfunctions $\phi_n(x)$ according to the following relation:
\begin{equation}
    \rho_{\textrm{LDOS}}(x,E)=\frac{1}{\pi}\sum_n |\phi_n(x)|^2 \Im \frac{1}{E-E_n-i\delta \epsilon}.
\end{equation}
$\delta \epsilon \approx 0.02eV$ is the approximate broadening of the energy spectrum peaks. The result is plotted in Fig.~\ref{fig:LDOS}, where the discrete energy levels of the system are clearly visible. The current measured here is then quantized with respect to $k_\theta$.  Therefore, the response is identified as a quantum spin-Hall current, which can be used to define the Hall conductivity. Through symmetry arguments it was proposed that the quantum Hall effect cannot exist on the M\"obius strip but the quantum spin-Hall effect should be possible \cite{Nontrivial}. This result is also confirmed here through this direct simulation.

 Experimental realisations of M\"obius strips of single crystals of NbSe$_3$ have been reported  \cite{Tanda2002} and recently a supra-molecular strategy for fabrication was proposed \cite{fabrication}.  Graphene M\"obius strips could be realized, by similar methods or in combination with graphene shape engineering tools such as optical forging \cite{opticalforging}. 
It can be shown that a pulsed laser beam can forge a graphene sheet into controlled three-dimensional shapes in the nanoscale, therefore two halves of a M\"obius ribbon can be forged, and further connected together \cite{selfassembly}. Additionally, due to the topological nature of the current the result will be robust to any possible temperature fluctuation. $\bar{J}^r_{P,H}$  can be measured across $r$ at a fixed $k$.





\subsection{Curvature-induced Hall current on the torus}
\begin{figure}[ht]
\includegraphics[width=\columnwidth]{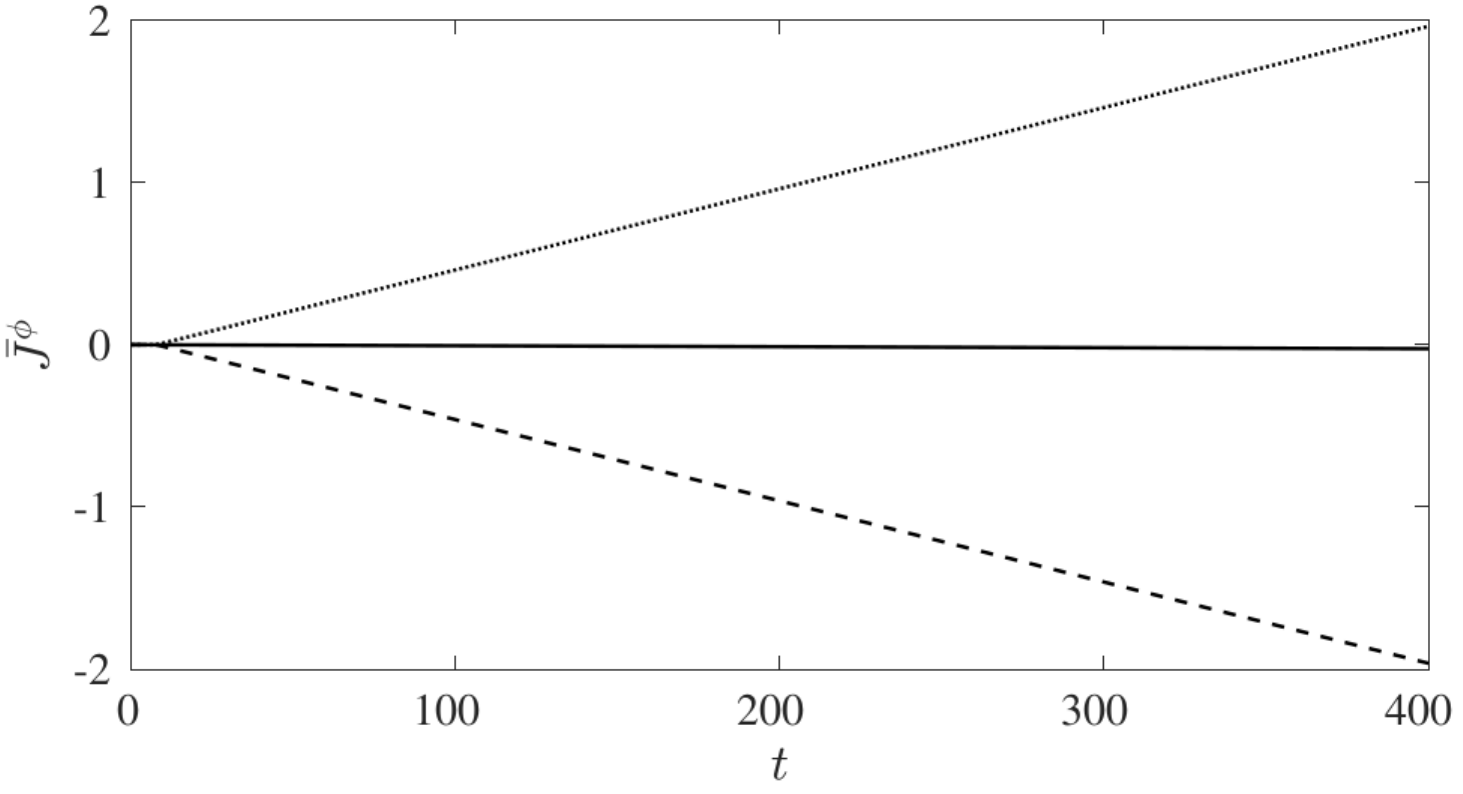}
\caption{\label{fig:torus_current} The time-evolution of the azimuthal current $\bar{J}^\phi$ for a non-zero momentum particle, (dotted line) and hole (dashed line) wavefunctions. The solid line represents a particle wavefunction with $k_\theta=0$. One simulational time-step, t, is equivalent to $1ps$. }
\end{figure}

To investigate further the relevance of topology and curvature, a similar but simpler shape is also simulated. The system is initialized to the torus geometry by the discrete mapping,
  \begin{align}
    h	=  & \begin{pmatrix}
           [R+ w  \cos(\phi)] cos(\theta) \\
           [R+ w  \cos(\phi)]sin(\theta) \\
           w  \sin(\theta)
         \end{pmatrix},
  \end{align}
with $\theta,\phi \in \{-\pi, \pi \}$,  width $w$, mid-circle of radius $R$, 
$R,w \in \mathbb{R}_{>0}$, both boundaries are periodic. In this parametrization the metric is diagonal and $\phi$ dependent $g_{ij}=\mathrm{diag}([R+w\cos(\phi])^2,w^2)$. In  Fig.~\ref{fig:torus_current} the time-evolution of the azimuthal current $\bar{J}^\phi$ is plotted for a particle, (dotted line) and hole (dashed line) wavefunctions as before. The solid line represents a particle wavefunction with $k_\theta=0$.  Here, it is evident that an equivalent Hall current develops for both particles and holes in the presence of a $\theta$ velocity. The different geometry results in a simpler behaviour, where a transverse current develops as a consequence of curvature. The net current, and spin-hall currents would be zero on a graphene torus.    This result clarifies that the spin-hall current observed in the previous section is of topological nature, developed from the introduction of the non-trivial topology, the M\"obius strip.

\section{Conclusions and Outlook}
We have presented a study of the topological and geometrical transport properties of a M\"obius graphene strip. The challenges and resolutions in simulating such a system were outlined. The continuity of the spin connection was achieved by a gauge transformation and limiting the half-width of the strip.


In the absence of a magnetic field, we measure a quantum spin-Hall current on the graphene strip originating from topology and curvature, whereas a quantum Hall current is not observed. This result also represents an example of the correspondence between a magnetic field and a curved manifold. The torus geometry is simulated for comparison, where a Hall current is measured. Additionally, a concrete illustration of the equivalence between the Berry and Ricci curvature is presented through a wave-packet traveling around a M\"obius strip. 

Building on these results, higher order and different types of topologies can be further investigated in the context of curvature. Without the need of a magnetic or electric field and by further understanding the properties of topological graphene sheets, simple and topologically robust quantum devices could be developed just by exploiting their geometry. 

\begin{acknowledgments}
The authors are grateful for the financial support Swiss National Science Foundation, under Grant No. 200021 165497 and Prof. S. Succi and I. Petrides for helpfull conversations.
\end{acknowledgments}

\bibliographystyle{ieeetr}
\bibliography{allcitations.bib}


\appendix
\section{Curved-Space Quantum Lattice Boltzmann \label{sec:QLBM}}

The quantum lattice Bolzmann (QLB) Method used for solving the Dirac equation as minimally coupled to curved space is an extension of the original method developed by Succi et al.  \cite{succi_qlbm}. The method exploits the conceptual similarities between the Dirac equation and the Boltzmann equation on the lattice.  
We present here the QLB method for a three-dimensional manifold.

\subsection{The Dirac Equation}

The Dirac equation can be naturally extended to curved space described by a metric tensor $g_{\mu\nu}$ with a covariant derivative $D_{\mu}$ as 
\begin{equation}
\label{eq:Dirac_cs}
(i \g^{\mu}D_{\mu}+m)\Psi=0,
\end{equation}
where $\g^{\mu}$ denote the Dirac matrices. 

The classical Boltzmann equation for a particle density distribution function $f(x_a, v_a, t)$ is given by:
\begin{equation}
\del_t f + v^i \del_{x^i}f=\mathcal{C}[f]-F^a\del_{v^a}f,
\end{equation}
 the left-hand side describes the advection of the distribution function, velocity $v^a$, whereas 
 the right-hand side describes the collisions between particles and the effect of external forces $F^a$. 
 Furthermore, the Dirac equation in curved space Eq.~\ref{eq:Dirac_cs} can be cast 
 into a kinetic theory form,
\begin{equation}
\label{eq:dirac_qlbm}
\del_t \Psi + \s^a\del_a \Psi = \mathcal{C} \Psi + \mathcal{F} \psi.
\end{equation}
Therefore, similarly to the Boltzmann equation,  the left-hand side represents the 
'free streaming' step along matrix valued 'velocities' $\s^i$ while the right-hand side 
contains a 'collision' and a 'forcing' term.

The collision term of Equation~(\ref{eq:dirac_qlbm}) is represented by:
\begin{equation}
\label{eq:collision}
\mathcal{C}=-(i m \g^0 + \s^a e_a^i\G_i), 
\end{equation}
where $m$ is the fermion mass.
The 'forcing term' is given by:
\begin{equation}
\label{eq:forcing_qlbm}
\mathcal{F}=-\s^a(e_a^i-\d_a^i) \del_i.
\end{equation}
where the symbols have their usual meaning. 
The partial derivative of the Dirac equation is distributed between the streaming part and the 
forcing term, resulting in a lattice-compatible classical streaming operator of 
the form $\del_t + v^a\del_a$, where $v^a \in \mathbb{Z}$. 
The forcing term is a consequence of the generalized Dirac matrices $\g^i=e_a^{~i}\g^a$ and captures the bulk of the curvature effects. The partial derivative in Equation~(\ref{eq:forcing_qlbm}) is approximated by a local finite difference scheme .    

\subsection{Diagonal Streaming Operator}

In order to obtain a diagonal streaming operator, the complex $\s$-matrices have to be diagonalized first, which 
yields a diagonal velocity matrix with eigenvalues $v^a=\pm 1$. 
The diagonalization is achieved by suitable "rotation matrices":

\begin{align*}
	X_a^\dagger \,\s^a\, X_a  
	= \begin{pmatrix}
			 1 & 0 & 0 & 0 \\
			 0 & 1 & 0 & 0 \\
			 0 & 0 & -1 & 0 \\
			 0 & 0 & 0 & -1
		\end{pmatrix}
	= \g^0 \qquad \text{for } a=0,1,2,
\end{align*}
where the unitary transformation matrices $X_1, X_2, X_3$ are given by:

\begin{align*}
   \setlength{\arraycolsep}{2pt}
   \renewcommand{\arraystretch}{0.8}
	X_1 &= \T\frac{1}{\sqrt 2} 
		\begin{pmatrix}
			 1 & 0 & -1 & 0 \\
			 0 & 1 & 0 & -1 \\
			 0 & 1 & 0 & 1 \\
			 1 & 0 & 1 & 0
		\end{pmatrix},
	~
	X_2 = \T\frac{1}{\sqrt 2} 
		\begin{pmatrix}
			 0 & i & 0 & 1 \\
			 -i & 0 & i & 0 \\
			 -1 & 0 & -1 & 0 \\
			 0 & -1 & 0 & -i
		\end{pmatrix},
		\\
        X_3 &= \T\frac{1}{\sqrt 2} 
		\begin{pmatrix}
			 1 & 0 & 0 & -1 \\
			 0 & 1 & 1 & 0 \\
			 1 & 0 & 0 & -1 \\
			 0 & 1 & 1 & 0
		\end{pmatrix}.
\end{align*}

The streaming and collision operations are performed in successive steps using operator splitting, since 
the simultaneous diagonalization of the three $\s$ matrices is not possible:
 \begin{align*}
	\Psi(t+\frac{\dt}{D}) &=
	\exp\big(-\dt\s^1\del_1+\frac{\dt}{D}(\mathcal{C}+\mathcal{F})\big) \Psi(t), \\
	\Psi(t+\frac{2\dt}{D}) &=
	\exp\big(-\dt\s^2\del_2+\frac{\dt}{D}(\mathcal{C}+\mathcal{F})\big) \Psi(t+\frac{\dt}{D}), \\
		\Psi(t+\dt) &=
	\exp\big(-\dt\s^3\del_3+\frac{\dt}{D}(\mathcal{C}+\mathcal{F})\big) \Psi(t+\frac{2\dt}{D}),
\end{align*}
where $D=3$ denotes the spatial dimensions. Each streaming step can be diagonalized by left multiplying with $X_a^\dagger$.
 \begin{align}
 \label{eq:diagonalization}
X_a^\dagger	\Psi(t+\frac{\dt}{D}) &=
	\exp\big(-\dt\s^a\del_a+\dt(\tilde{\mathcal{C}}_a+\tilde{\mathcal{F}}_a)\big) \tilde{\Psi}_a(t),
\end{align}
with the definitions:
 \begin{align*}
\Tilde{\Psi}_a \coloneqq X_a^\dagger \Psi, ~~ \tilde{\mathcal{F}}_a \coloneqq \frac{1}{2}X_a^\dagger \mathcal{F} X_a, ~~ \tilde{\mathcal{C}}_a \coloneqq \frac{1}{2} X_a^\dagger \mathcal{C} X_a,
\end{align*}
for $a=1,2,3$ (no Einstein summation is used here).
The exponential is approximated as:
 \begin{align*}
& \exp\big(-\dt\s^a\del_a+ \dt(\tilde{\mathcal{C}}+\tilde{\mathcal{F}})\big) \\
& \approx \big(\mathbb{I} -\dt\s^a \del_a+\dt(\tilde{\mathcal{C}_a}+\dt\tilde{\mathcal{F}}_a)+
\\
&+(\mathbb{I} - \frac{\dt}{2}\tilde{\mathcal{C}}_a)^{-1}(\mathbb{I} +\frac{\dt}{2}\tilde{\mathcal{C}}_a)\big)
\end{align*}
The expansion of the collision operator $e^{\dt\tilde{\mathcal{C}}_a}$ is unitary and thus conserves exactly the probability of the wavefunction. The streaming $e^{-\dt\g^0\del_a}$ and forcing $e^{\dt\tilde{\mathcal{F}}_a}$ operators are not expanded, as this is prohibited by the derivative. A simple $2^{nd}$-order expansion is performed, limiting the probability norm to $\Delta t^2$ accuracy. 
The operator splitting implies an error of order $\mathcal{O}(\dt^2)$, as 
$e^{\dt X}\cdot e^{\dt Y}=e^{\dt(X+Y)+1/2\dt^2[X,Y]}=e^{\dt(X+Y)}+\mathcal{O}(\dt^2)$.

The manifold is described by a chart $h$ defined in linear space, discretized on a regular rectangular lattice. 
The curved space Quantum Lattice Boltzmann Method evolves the four-spinor $\Psi = (\Psi^+, \Psi^-) = (\Psi_1^+,\Psi_2^-,\Psi_1^-,\Psi_2^+)$ from $t$ to $t+\delta t$. 
Once the operators are split, the following algorithm is performed in sequence for each 
lattice direction $n_a$, where $n_1=(1,0)$, $n_2=(0,1)$, and $a=1,2$. 

\begin{enumerate}
\item \textbf{Rotation:} The spinor is rotated by $X_a$, 
\begin{equation}
\tilde{\Psi}_a(x,t)=X^\dagger_a \Psi(x,t).
\end{equation}

\item \textbf{Collisions and curvature:} The collision and force operators are applied to the rotated spinor,
\begin{equation*}
\tilde{\Psi}_a^*(x,t)=\big( \Delta t \tilde{\mathcal{F}}_a + ( \mathbb{I} - \frac{\Delta t}{2} \tilde{\mathcal{C}}_a)^{-1}  ( \mathbb{I} + \frac{\Delta t}{2} \tilde{\mathcal{C}}_a)   \big) \tilde{\Psi}_a(x,t),
\end{equation*}
 where $\mathbb{I}$ is the identity matrix and $\tilde{\Psi}_a^*(x,t)$ denotes an auxiliary field,
\begin{equation}
\label{eq:collisiontilde}
\tilde{\mathcal{C}}_a=\frac{1}{2}X^\dagger_a \mathcal{C}X_a=-\frac{i}{D}m(X^\dagger_a \gamma^0 X_a)- \gamma^0e_a^i\Gamma_i,
\end{equation}

\begin{equation}
\label{eq:forcingtilde}
\tilde{\mathcal{F}}_a\tilde{\Psi}_a(x,t)=\big(e_a^i-\delta_a^i\big) \Big(\tilde{\Psi}_a(x \mp n_i \Delta t,t)-\tilde{\Psi}_a(x,t) \Big),
\end{equation}

where $n_i$ is the lattice direction and $\mathcal{C}$ is the collision term, Equation~(\ref{eq:collision}).
The upper sign applies to the spin-up components $(\Psi_1^+,\Psi_2^+)$ and the lower sign to the spin-down 
components $(\Psi_1^-,\Psi_2^-)$.
\item \textbf{Streaming:} The spinor components are streamed to the closest grid points along the lattice direction $\pm n_a$,
\begin{equation}
\tilde{\Psi}_a(x,t+\frac{\Delta t}{2})=\tilde{\Psi}_a^*(x \mp n_a\Delta t,t).
\end{equation}

\item \textbf{Inverse Rotation:} The spinor is rotated back via $X_a$,
\begin{equation}
\Psi_a(x,t+\frac{\Delta t}{2})=X_a \tilde{\Psi}_a(x,t+\frac{\Delta t}{2}).
\end{equation}

\item Repeat steps 2--4 for the next spatial direction.
\end{enumerate}

 The external potentials $V(x)$, scalar, and $A(x)$, vector are added to the collision operator Equation~(\ref{eq:collisiontilde}), such that:
\begin{equation}
\tilde{\mathcal{C}}_a=\frac{1}{2}X^\dagger_a \mathcal{C}X_a=-\frac{i}{D}(m-V)(X^\dagger_a \gamma^0 X_a)- \gamma^0e_a^i(\Gamma_i-iA_i).
\end{equation}
 
The simulation for strained graphene is carried out with modified Equations~(\ref{eq:collisiontilde}) and (\ref{eq:forcingtilde}), according
to the following scheme:  
\begin{equation*}
\tilde{\mathcal{C}}_a \rightarrow \sqrt{g} \tilde{\mathcal{C}}_a, ~ e_a^i\rightarrow \sqrt{g}e_a^i.
\end{equation*}
The additional factor $\sqrt{g}$ originates from the volume element of the Hamiltonian Equation~(\ref{eq:hamiltoniangraphene}).

\section{Riemannian geometry}
\label{app:riemannian}

The Latin indices run over the spatial dimensions and Einstein summation convection is used for repeated indices.

A $D$ dimensional curved space is represented by a Riemannian manifold M, which is locally described by a smooth diffeomorphism $\mathbf{h}$, called the chart. The set of tangential vectors attached to each point $\mathbf{y}$ on the manifold is called the  tangent space $T_\mathbf{y} M$. In the fluid model, all the vector quantities are represented as elements of $T_\mathbf{y} M$. The derivatives of the chart $\mathbf{h}$ are used to define the standard basis $(\textbf{e}_1,...,\textbf{e}_D)=\frac{\del\mathbf{h}}{\del x^1},...,\frac{\del \mathbf{h}}{\del x^D}$. 

The metric tensor $g$ can be used to measure the length of a vector or the angle between two vectors. In local coordinates, the components of the metric tensor are given by 
\begin{equation}
g_{ij}(x)= \textbf{e}_i(x)\cdot \textbf{e}_j(x)= \frac{\del \mathbf{h}}{\del x^i} \cdot \frac{\del \mathbf{h}}{\del x^j},
\end{equation}
where $\cdot$ is the standard Euclidean scalar product.

For a given metric tensor, the vector $v=v^i\textbf{e}_i \in T_\mathbf{y} M$ has a norm $||v||_g=\sqrt{v^ig_{ij}v^j}$ and a corresponding dual vector $v^*=v^i\textbf{e}_i \in T^*_\mathbf{y} M$ in the cotangent space, which is spanned by the differential 1-forms $dx^i=g(\textbf{e}_i,\cdot)$. The coefficients $v_i$ of the dual vector are typically denoted by a lower index and are related to the upper-index coefficients $v^i$ by contraction with the metric tensor $v_i = g_{ij}v^j$ or equivalently, $ v^i=g^{ij}v_j$, where $g^{ij}$ denotes the inverse of the metric tensor. The upper-index coefficients $v^i$ of a vector $v$ are typically called  \textit{contravariant components}, whereas the lower-index coefficients $v_i$ of the dual vectors $v^*$ are known as the \textit{covariant components}.

A necessary feature for the description of objects moving on the manifold is parallel transport of vectors along the manifold.  The tangent space is equipped with a covariant derivative $\nabla$ (Levi-Civita connection), which connects the tangent spaces at different points on the manifold and thus allows to transport a tangent vector  from one tangent space to the other along a given curve  $\gamma(t)$. The covariant derivative can be viewed  as the orthogonal projection of the Euclidean derivative $\del$ onto the tangent space, such that the tangency of the vectors is preserved during the transport. In local coordinates, the covariant derivative is fully characterized by its connection coefficients $\Gamma^i_{jk}$  (Christoffel symbols), which are defined by the action of the covariant derivative on the basis vector, $\nabla_j \textbf{e}_k= \Gamma^i_{jk}$. In the standard basis, $\textbf{e}_i = \frac{\del \mathbf{h}}{\del x^i}$, the Christoffel symbols are related to the metric by
\begin{equation}
\Gamma^i_{jk}=\frac{1}{2}g^{ij}(\del_j g_{kl} + \del_k g_{jl} - \del_l g_{jk}).
\end{equation} 
Acting on a general vector $v=v^i \mathbf{e}_i,$ the covariant derivative becomes:
 \begin{equation}
\nabla_k v =(\del_k v^i + \Gamma^i_{kj}v^j)\mathbf{e}_i,
\end{equation}
where the product rule has been applied, using that the covariant derivative acts as a normal derivative on the scalar functions 
$v^i$. Extending to tensors of higher rank, for example the second order tensors $T= T^{ij} $, 
\begin{equation}
\nabla_k T=( \del_kT^{ij}+ \G ^i _{kl} T^{lj} + \G^j_{kl} T^{il})\mathbf{e}_i \otimes \mathbf{e}_j
\end{equation}
in this work the basis vectors $\mathbf{e}_i $ are generally dropped. Compatibility of the covariant derivative with the metric tensor implies that $\nabla_k g^{ij}=\nabla_k g_{ij} =0$. This property allows us to commute the covariant derivative with the metric tensor for the raising or lowering of tensor indices in derivative expressions.

The motion of the particle can be described by the curve $ \gamma(t)$, which parametrizes the position of the particle at time $t$. The geodesic equation,  $ \nabla_{\dot{\gamma}} \dot{\gamma} =0 $, in local coordinates $\gamma(t)=\gamma^i(t)\mathbf{e}_i$ is defined by
\begin{equation}
\label{eq:geodesic}
\ddot{\gamma}^i + \Gamma_{jk}^i \dot{\gamma^j} \dot{\gamma^k} = 0. 
\end{equation}
The geodesic equation can be interpreted as the generalization of Newtons law of inertia to curved space. The solutions of Eq.~(\ref{eq:geodesic}) represent lines of constant kinetic energy on the manifold, i.e. the geodesics. 
	The Riemann curvature tensor $R$ can be used to measure curvature, or more precisely, it measures curvature-induced change of a tangent vector $v$ when transported along a closed loop.
 \begin{equation}
R(\textbf{e}_i,\textbf{e}_j)v=\nabla_i \nabla_j v-\nabla_j \nabla_i v.  
\end{equation}   
In a local coordinate basis $ { \textbf{e}_i } $, the coefficients of the Riemann curvature tensor are given by
\begin{align}
R^l_{ijk}&= g(R(\textbf{e}_i,\textbf{e}_j)\textbf{e}_k,\textbf{e}_l) 
\\
&=\del_j \Gamma^l_{ik} - \del_k \Gamma^l_{ij} + \Gamma^l_{jm} \Gamma^m_{ik} -\Gamma^l_{km} \Gamma^m_{ij}.
\end{align}
Contraction of $R^i_{jkl}$ to a rank 2 and 1 tensor yields  the Ricci-tensor $R_{ij}=R^k_{ikj}$ and the Ricci-scalar $R=g^{ij}R_{ij}$ respectively, which can also be used to quantify curvature.

The gradient is defined as $\nabla^i f= g^{ij} \del_j f$, the divergence as $\nabla_i v^i= \frac{1}{\sqrt{g}} \del_i (\sqrt{g} v^i) $, and the integration over curved volume as $V=\int_V QdV$, where $dV=\sqrt{g}dx^1...dx^D=:\sqrt{g}d^Dx$ denotes the volume element. $\sqrt{g}$ denotes the square root of the determinant of the metric tensor.   

It should be clarified that in the simulations there is no time curvature and $g_{ij}$ denotes the curved space metric.

\section{Convergence of the closed Dirac boundary conditions \label{app:boundary}}
\begin{figure}[htbp]
\includegraphics[width=1\columnwidth]{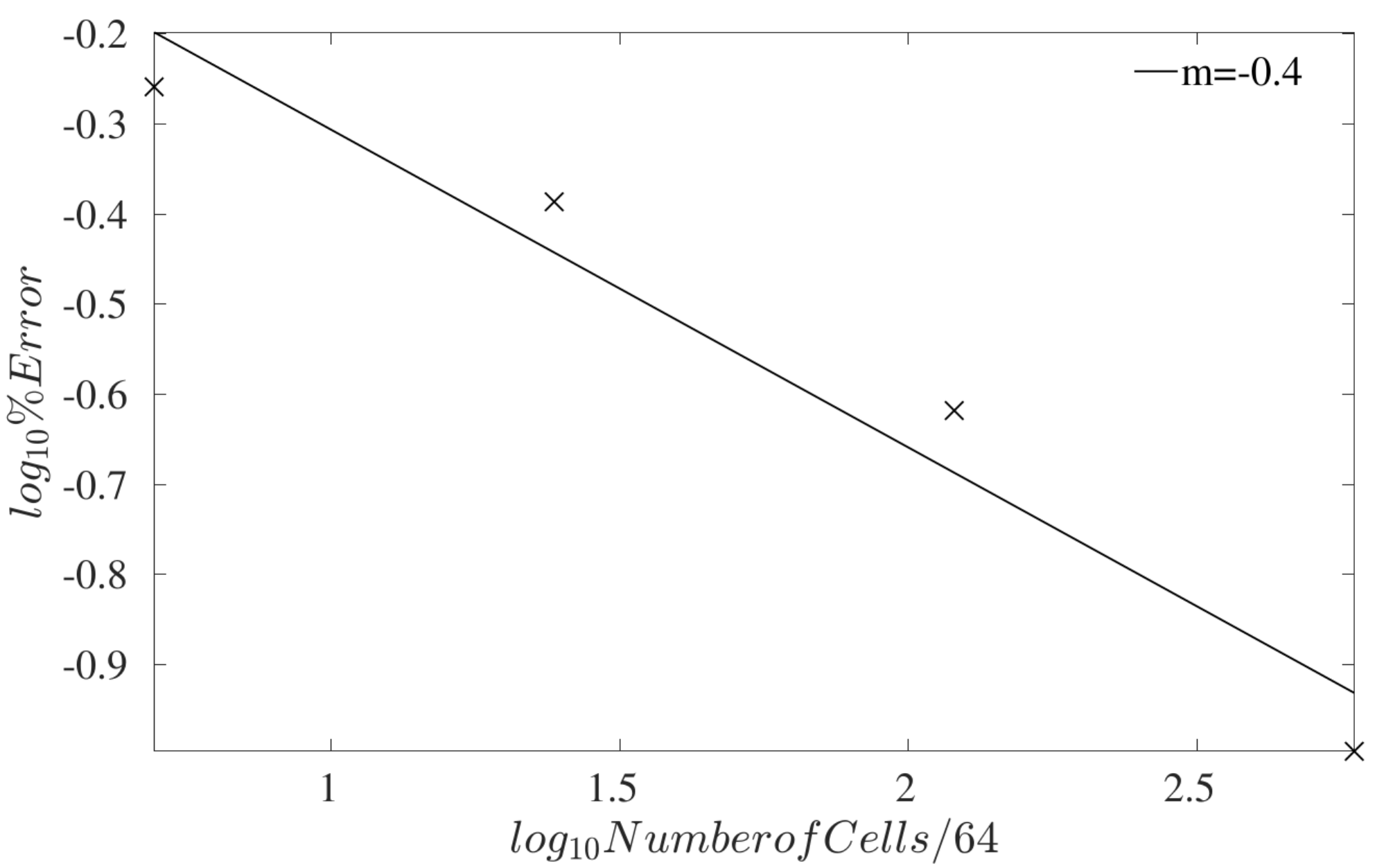}
\caption{\label{fig:convergence} Error convergence of the proposed quantum lattice Boltzmann closed boundary condition. The logarithmic of the relative error of the density $\rho$ is plotted against the normalised logarithmic number of cells. $m$ denotes the gradient. }
\end{figure}

The convergence of the proposed quantum lattice Boltzmann boundary condition is shown on Figure \ref{fig:convergence}.

\section{Ricci scalar for the M\"obius strip  \label{app:ricci_moebius}}
\begin{figure}[htbp]
\includegraphics[width=1\columnwidth]{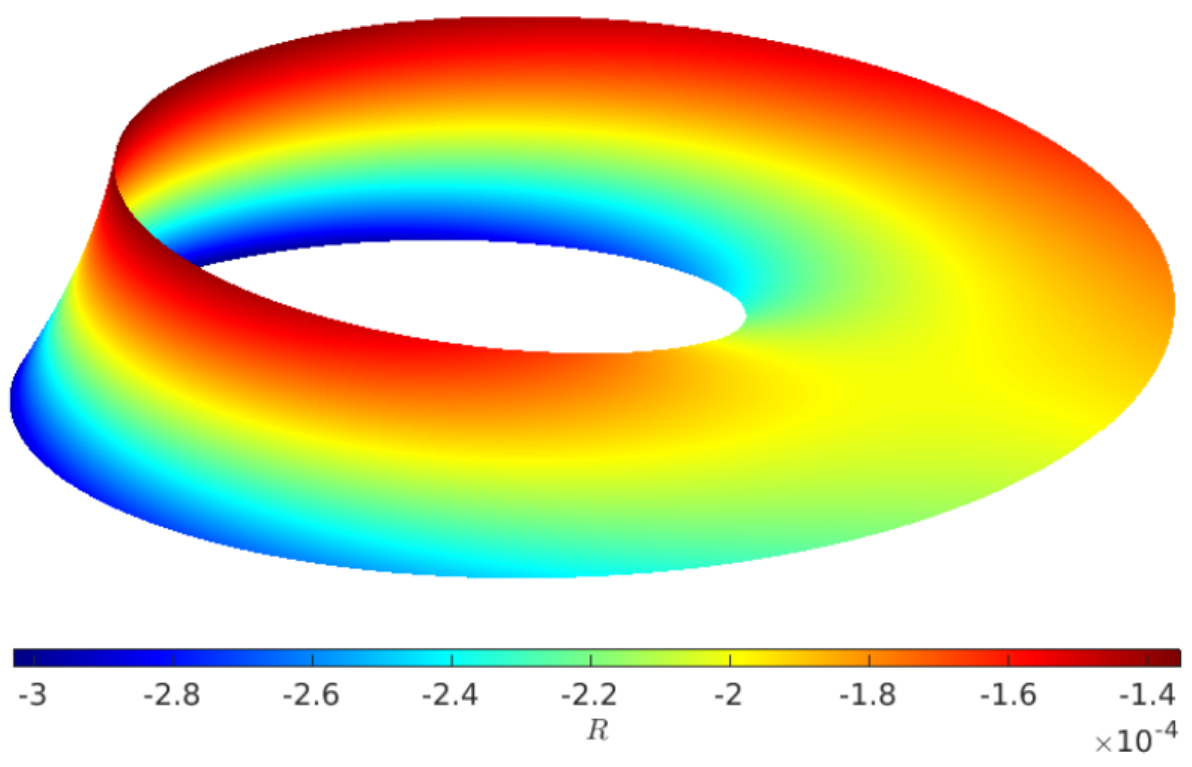}
\caption{\label{fig:ricci} The Ricci scalar of the M\"obius strip for half-width $w=0.1$ and mid-circle of radius $R=1$.  }
\end{figure}

The Ricci scalar for the M\"obius strip is non-zero across the whole domain for the specific choice of parameters as shown on Fig.~\ref{fig:ricci}.

\end{document}